\documentstyle[multicol,aps,epsf]{revtex}
\begin{document}

\title{A note on dissipation in helical turbulence}

\author{
P. D. Ditlevsen and P. Giuliani\\
The Niels Bohr Institute,
University of Copenhagen,\\ Blegdamsvej 17,
 DK-2100 Copenhagen O, Denmark.}
\date{\today}
\maketitle
\begin{abstract}
In helical turbulence 
a linear cascade of helicity accompanying the energy cascade 
has been suggested. Since
energy and helicity have different dimensionality
we suggest the existence of a characteristic inner scale,  $\xi=k_H^{-1}$, 
for helicity
dissipation in a regime of hydrodynamic fully developed turbulence
and estimate
it on dimensional grounds. This scale is always larger than the
Kolmogorov scale, $\eta=k_E^{-1}$, 
and their ratio $\eta / \xi $ vanishes in the high Reynolds
number limit,  so the flow
will always be helicity free in the small scales.
\end{abstract}
\begin{multicols}{2}
In helical turbulence coexisting cascades of energy and helicity
was envisaged by Brissaud et al. \cite{Brissaud}. Based on 
dimensional analysis it was conjectured that the helicity 
cascade is linear in the sense that the spectral helicity
density follows the spectral energy density, $H(k)\propto E(k)
\propto k^{-5/3}$. This scenario was supported numerically 
by Andr{\'e} and Lesieur in an EDQNM closure calculation \cite{Andre}
and by Borue and Orzag \cite{Borue} in a direct numerical 
simulation. Following Brissaud et al. the existence of a 
linear helicity cascade is due to an equal distortion time
leading to the non-linear transfer of energy and helicity. 
The distortion time at a scale $k$ is estimated as \cite{Kraichnan},

\begin{equation}
\tau_{k}\sim (\int_0^k p^2E(p)dp)^{-1/2}. 
\label{tau}
\end{equation}
Here and in the following $\sim$ denotes 'equal within
order unity constants' \cite{Frisch}. 
The non-linear transfers of energy and helicity are then,

\begin{equation}
\Pi_E(k) \sim kE(k)/\tau(k)
\label{PiE}
\end{equation}
and
\begin{equation}
\Pi_H(k) \sim kH(k)/\tau(k).
\label{PiH}
\end{equation}
From (\ref{tau}) and (\ref{PiE}) the K41 result,

\begin{equation}
E(k)\sim \overline{\varepsilon}^{2/3}k^{-5/3},
\label{ek}
\end{equation}
follows where $\overline{\varepsilon}$ is the mean
energy dissipation or mean non-linear energy transfer or mean energy
input. 
Correspondingly, from (\ref{tau}) and (\ref{PiH}) we obtain, 
\begin{equation}
H(k)\sim \overline{\delta}\overline{\varepsilon}^{-1/3}k^{-5/3},
\label{hk}
\end{equation}
where $\overline{\delta}$ is the mean helicity input.
The linear helicity cascade is derived under the assumption 
that helicity dissipation is negligible in the inertial range. 
The helicity
density is $h=u_i\omega_i$/2, where $\omega_i=\epsilon_{ijk}\partial_ju_k$ is
the vorticity. Conventionally the helicity is defined as
$2h$, this is not important for the discussion presented here.
An instructive way of representing this spectrally is to expand the
velocity vector $u_i({\bf k})$ in a basis
of 'helical modes' \cite{waleffe;1992}.
The helical modes ${\bf h}_\pm$ are the (complex) eigenvectors
of the curl operator,
$i {\bf k} \times {\bf h}_\pm = \pm k {\bf h}_\pm$.
Using incompressibility, ${\bf k}\cdot {\bf u}({\bf k})=0$,
we have ${\bf u}({\bf k})=u_+({\bf k}){\bf h}_+ + u_-({\bf k}){\bf h}_-$ and the energy
and helicity in the mode ${\bf u}({\bf k})$ are,
\begin{equation}
E({\bf k})={\bf u}({\bf k})\cdot {\bf u}({\bf k})^*/2=(|u_+({\bf k})|^2+|u_-({\bf k})|^2)/2
\end{equation}
and

\begin{equation}
H({\bf k})={\bf u}({\bf k})\cdot {\bf \omega}({\bf k})^*/2=k(|u_+({\bf k})|^2-|u_-({\bf k})|^2)/2.
\end{equation}

The spectral energy and helicity densities can then be separated 
into the densities
of modes of positive and negative helicity $E(k)=E_+(k)+E_-(k)$
and $H(k)=H_+(k)+H_{-}(k)=k(E_+(k)-E_-(k))$. 
From this we have the rigorous
constraint on the spectral helicity density,

\begin{equation}
|H(k)|\le kE(k).
\label{hke}
\end{equation}
A similar constraint can be derived regarding the mean inputs
of energy $\overline{\varepsilon}$ and helicity $\overline{\delta}$.
Suppose the flow is forced with a forcing ${\bf f}$ at the pumping
scale such that ${\bf f(k)}=0$ for $|{\bf k}| > K$ where $K$ 
is a wavenumber larger than the pumping scale. Then it follows
that 
$|\overline{\delta}|\le K\overline{\varepsilon}$ \cite{Borue}, 
where $K$ is a wavenumber at the pumping scale. 
When the scaling relations (\ref{ek}) and (\ref{hk}) are applied
to the densities of positive and negative helicities
separately, there
must be a detailed cancellation of the leading scaling, such 
that,

\begin{equation}
E_+(k)=(C/2) \overline{\varepsilon}^{2/3}k^{-5/3}+(C_H/2)
\overline{\delta}\overline{\varepsilon}^{-1/3}k^{-8/3}
\end{equation}
and
\begin{equation}
E_-(k)=(C/2) \overline{\varepsilon}^{2/3}k^{-5/3}-(C_H/2)
\overline{\delta}\overline{\varepsilon}^{-1/3}k^{-8/3}
\end{equation}
where $C$ and $C_H$ are some (non-universal) order unity
Kolmogorov constants.

The energy dissipation is given as, $D_E=\nu \int_0^{k_E}
k^2 E(k)dk$, and the upper limit of the integral which
is the (inverse) Kolmogorov scale $k_E$
is as usual determined
by $\nu k_E^3 E(k_E) \sim \nu k_E^3 
(\overline{\varepsilon}^{2/3}k_E^{-5/3}) \sim \overline{\varepsilon} 
\Rightarrow k_E \sim (\overline{\varepsilon}/\nu^3)^{1/4}$.
The dissipation is linear and can thus be split into
dissipation of the positive and negative helicity
parts of the spectrum separately. This implies that the
dissipation of one sign of helicity $(s=\pm)$ is
$D_{H_s}\sim \nu \int_0^{k_H}k^2 H_s(k)dk=\nu k_H^4 E_s(k_H)$. The
helicity of sign $s$ is thus dissipated at a scale determined by,

\begin{eqnarray}
D_{H_s}\sim\nu k_H^4 E_s(k_H)\\ \nonumber \sim\nu k_H^4(\overline{\varepsilon}^{2/3}k_H^{-5/3}
+s\overline{\delta}\overline{\varepsilon}^{-1/3}k_H^{-8/3})\sim\overline{\delta}
\end{eqnarray}
and we
arrive at an (inverse) inner scale $k_H$, different from the Kolmogorov scale $k_E$,
for dissipation of helicity,

\begin{equation}
k_H \sim [\overline{\delta}^3/(\nu^3\overline{\varepsilon}^2)]^{1/7}.
\label{kH}
\end{equation}
Note that this scale can not be obtained by pure dimensional counting in
a manner similar to the Kolmogorov scale $k_E \sim 
\overline{\varepsilon}^\alpha \nu^\beta \Rightarrow 
(\alpha,\beta)=(1/4,3/4)$. In the case of helical turbulence
we can define an (integral) length scale $L=(\overline{\varepsilon}/\overline{\delta})$ and thereby 
$k_H \sim \overline{\varepsilon}^\alpha \nu^\beta 
(k_H\overline{\varepsilon}/\overline{\delta})^\gamma$ from which
$\gamma$ is undetermined by dimensional counting. For $\gamma=0$
the Kolmogorov scale is obtained and for $\gamma=-3/4$ equation (\ref{kH}) is
obtained.

It is easy to see that for any flow realization we must have
$k_H \le k_E$, so a pure helicity cascade is not possible. 
This result can also be obtained by 
estimating where the flow should be forced in order
to dissipate the helicity at the Kolmogorov scale such that
$k_H\sim k_E$. Pumping helicity into the flow at wave number $\kappa$
implies $\overline{\delta} \sim \kappa \overline{\varepsilon}$.
We thus have,

\begin{equation}
k_H\sim k_E \Rightarrow
[\frac{(\kappa\overline{\varepsilon})^3}{\nu^3  \overline{\varepsilon}^2}]^{1/7}\sim
(\frac{\overline{\varepsilon}}{\nu^3})^{1/4} \Rightarrow 
\kappa \sim k_E.
\end{equation}
This shows that the the flow must be forced at the 
Kolmogorov scale which is in conflict with the 
existence of an inertial range. A similar result was
obtained by P. Olla \cite{Olla} in a different
way 
using an argument based on
the EDQNM approximation.

Furthermore,
we have  $k_H / k_E \propto \nu^{-3/7+3/4}=\nu^{9/28} \rightarrow 0$
for $\nu\rightarrow 0$. So again for high Reynolds number helical flow 
the small scales will always
be non-helical.
The inner scale for helicity dissipation plays a different 
role in helical turbulence than the Kolmogorov scale. The 
dissipation of one sign of helicity at a given wavenumber will grow with
wavenumber as $D_{H_s}(k) \propto k^{7/3}$, thus the dissipation
of either sign of helicity will grow with wavenumber in the 
range $k_H<k<k_E$. This is only possible if there is
a detailed balance between dissipation of positive and negative 
helicities in that range. 

In conclusion, the scenario we propose for high
Reynolds number helical turbulence is then the following. 
At the integral scale $K$ energy and helicity is forced into the
flow. in the inertial range $K<k<k_H$ there is a coexisting
cascade of energy and helicity where helicity follows a 'linear
cascade' with a $H(k) \sim k^{-5/3}$ spectrum. In the
range $k_H < k < k_E$ the dissipation of helicity dominates
with a detailed balance between dissipation of positive and
negative helicities and the right-left symmetry of the flow is
restored. The balanced positive and negative helicities
are generated in analogy to the enstrophy being generated
in high Reynolds number flow. 
The proposed scenario has been illustrated in a
shell model of turbulence \cite{ditlevsen;2001}. However, since the 
considerations presented here are purely phenomenological
they should be tested in experiments or numerical simulations.

%
%

\end{multicols}
\end{document}